\documentclass[10pt,twocolumn]{article}

\usepackage[T1]{fontenc}
\usepackage[utf8]{inputenc}
\usepackage{lmodern}
\usepackage[margin=1in]{geometry}
\usepackage{microtype}
\usepackage{setspace}
\usepackage{parskip}

\usepackage{amsmath,amssymb,amsthm,bm}

\usepackage{graphicx}
\usepackage{booktabs}
\usepackage{caption}
\usepackage{subcaption}
\usepackage{float}
\usepackage{rotating}
\usepackage{multirow}
\usepackage{array}

\usepackage[authoryear]{natbib}

\usepackage[colorlinks=true,linkcolor=blue,citecolor=blue,urlcolor=blue]{hyperref}

\usepackage{titlesec}
\titleformat{\section}{\large\bfseries}{\thesection.}{0.5em}{}
\titleformat{\subsection}{\normalsize\bfseries}{\thesubsection.}{0.5em}{}
\titleformat{\subsubsection}{\normalsize\itshape}{\thesubsubsection.}{0.5em}{}

\title{\vspace{-1cm}\textbf{Galactic Rotation Curves from Full-Disk Newtonian Gravity: The Lost and Found Model}}

\author{
Adolfo Santa Fe Dueñas\\[0.3em]
\small Space Science Center, University of New Hampshire, Durham, NH 03824, USA\\
\small Email: \texttt{adolfo.santafeduenas@my.utsa.edu}
}

\date{\today}

\begin{document}

\maketitle

\begin{abstract}
The approximately flat outer parts of spiral galaxy rotation curves are commonly interpreted as evidence for a discrepancy between the observed baryonic mass and the dynamical mass inferred from the measured orbital velocities. In many analyses, simplified mass estimates are often expressed through the relation $v^2(R)=GM(<R)/R$, which is exact only under spherical symmetry. Spiral galaxies, however, are flattened disk systems, for which mass exterior to the galactocentric radius under consideration can contribute non-negligibly to the gravitational field.

We introduce the \textit{Lost and Found} (LF) model, a geometrically consistent Newtonian framework based on direct full-disk gravitational integration and a parametrized representation of the disk surface density. This approach is closely related to previous thin-disk treatments that compute the gravitational field from the full mass distribution, while providing a simplified parametrization suitable for systematic fitting across heterogeneous galaxy samples.

We apply the LF model to a heterogeneous sample of disk galaxies spanning a broad range of masses and radial extents. The model reproduces the main observed features of the rotation curves, including the inner rise and the approximately flat outer behavior, while yielding systematically lower inferred masses compared to conventional dynamical mass estimates. Across the sample, the LF-inferred mass scales nearly linearly with the conventional dynamical mass, with a characteristic scaling factor $\eta_{\rm LF}\sim0.67$.

These results suggest that part of the inferred mass discrepancy in disk galaxies may be associated with geometric assumptions in standard mass estimates, and highlight the importance of full-disk treatments when interpreting galactic rotation curves.
\end{abstract}

\vspace{0.5em}

%==================================================
\section{Introduction}
\label{sec:introduction}

Since the mid-20th century, galactic rotation curves have played a central role in the study of mass distribution in galaxies. 

Observations of spiral systems, beginning with the pioneering work of \citet{rubin1980rotational},
showed that the rotational velocity of stars and gas does not exhibit the Keplerian decline ($v \propto r^{-1/2}$) expected for a centrally concentrated mass distribution, but instead remains approximately constant at large galactocentric distances.

This behavior has traditionally been interpreted as evidence for an additional, non-luminous mass component, commonly described in terms of dark matter halos \citep{navarro1996structure,burkert1995structure}. Alternative approaches, most notably Modified Newtonian Dynamics (MOND), have instead proposed modifications to the law of gravity or inertia at low accelerations \citep{milgrom1983modification}. More recently, empirical relations such as the radial acceleration relation (RAR) have highlighted the close connection between the observed dynamics and the visible mass distribution \citep{mcgaugh2016radial}.

Despite these different interpretations, many rotation-curve analyses share a common underlying step: the inference of gravitational acceleration from an assumed mass distribution. The circular velocity is often expressed through the relation
\begin{equation}
v(r)=\sqrt{\frac{G M(<r)}{r}},
\end{equation}
where $M(<r)$ is interpreted as the mass enclosed within radius $r$. This expression is exact under spherical symmetry, where Newton's shell theorem applies \citep{binney2011galactic}. Spiral galaxies, however, are not spherical systems, but flattened disk-like structures \citep{freeman1970disks}.

For disk geometries, the gravitational contribution of material exterior to the galactocentric radius under consideration does not generally vanish. Instead, the total acceleration must be obtained from the full mass distribution of the disk \citep{casertano1983rotation,binney2011galactic}. This suggests that the inferred relation between luminous and dynamical mass may depend not only on the amount of matter present, but also on how the gravitational field is evaluated in disk geometries.

Several previous studies have shown that Newtonian disk gravity can produce richer dynamical behavior than is captured by simplified enclosed-mass arguments \citep{casertano1983rotation,feng2012mass,jalocha2010}. The present work revisits this issue by focusing specifically on the contribution of the full disk mass distribution, including material exterior to the radius of interest, and by quantifying the resulting impact on inferred galactic masses across a heterogeneous sample of spiral galaxies.

To this end, we introduce the \textit{Lost and Found} (LF) model, a geometrically consistent Newtonian framework that combines direct full-disk gravitational integration with a two-component exponential representation of the disk surface density. This approach incorporates both the geometric contribution of the full disk and the radial structure of the mass distribution in a self-consistent manner, without imposing spherical symmetry. We show that it reproduces the main observed features of galactic rotation curves while systematically reducing the inferred total mass relative to conventional enclosed-mass estimates.

The structure of this paper is as follows. Section~\ref{sec:traditional} discusses the geometric limitations of the traditional enclosed-mass approach. Section~\ref{sec:lfmodel} introduces the LF model and its mathematical formulation. Section~\ref{sec:methodology} describes the numerical methodology. Section~\ref{sec:results} presents the rotation-curve fits and mass comparisons. Section~\ref{sec:discussion} discusses the physical interpretation and limitations of the model, and Section~\ref{sec:conclusions} summarizes the main conclusions.

\section{Geometric Considerations in Disk Gravity}
\label{sec:traditional}

\subsection{Geometric Context of the Enclosed-Mass Approximation}

A commonly used approximation in galactic dynamics relates the circular velocity to an enclosed mass through the enclosed-mass relation:

\begin{equation}
    v(r) = \sqrt{\frac{G M(<r)}{r}},
\end{equation}

where $M(<r)$ is interpreted as the mass enclosed within radius $r$. This relation is exact under spherical symmetry, where Newton's shell theorem applies \citep{Newton1687,binney2011galactic}.

Under spherical symmetry, the gravitational contribution from mass located outside the radius of interest cancels exactly. This cancellation arises from the balance between the increase in projected area ($\propto r^2$) and the decrease in gravitational force ($\propto 1/r^2$), leading to a net zero contribution from spherical shells.

\begin{figure}
    \centering
    \includegraphics[width=0.4\textwidth]{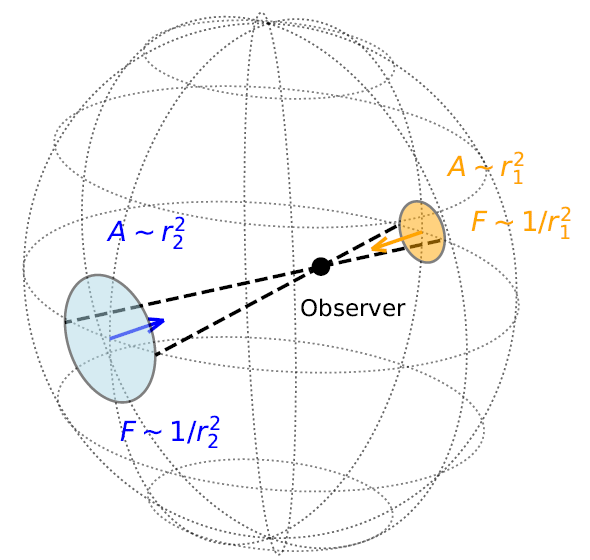}
    \caption{
    Gravitational force cancellation in a spherical shell. The projected area of distant mass elements increases as $r^2$, while the gravitational force decreases as $1/r^2$, leading to exact cancellation. This result depends on spherical symmetry. Adapted from \protect\cite{binney2011galactic}.
    }
    \label{fig:shell_theorem}
\end{figure}

This behavior is illustrated in Figure~\ref{fig:shell_theorem}.

Spiral galaxies, however, are not spherical systems, but rather flattened, disk-like structures \citep{freeman1970disks}. In such geometries, the conditions required for exact cancellation are not satisfied.

As a result, material located at radii larger than the observation point can contribute non-negligibly to the total gravitational field. In this case, the acceleration is determined by the full mass distribution of the disk rather than by the enclosed mass alone \citep{casertano1983rotation,binney2011galactic}.

\subsection{Illustrative Geometric Considerations in Disk Systems}

To provide an intuitive geometric illustration of this effect, we consider a simplified disk model with uniform surface density and total radius $R_{\mathrm{total}}$.

This construction is not intended as a quantitative description of disk gravity, but rather as a simple geometric illustration of how material exterior to the observation radius may contribute to the net gravitational field in disk systems.

In the enclosed-mass approximation, the gravitational field at a radius $R_{\mathrm{obs}}$ is estimated using only the mass within that radius, corresponding to a circular region centered on the galactic nucleus:

\begin{equation}
A_{\mathrm{trad}} = \pi R_{\mathrm{obs}}^2.
\end{equation}

In a disk geometry, however, the contribution from material exterior to $R_{\mathrm{obs}}$ does not generally cancel. For the special case of uniform surface density, one may construct a geometrically symmetric region centered on the observer whose contributions tend to cancel locally, providing a simple intuitive picture of how different parts of the disk contribute to the net gravitational field.

\begin{figure}
    \centering
    \includegraphics[width=0.5\textwidth]{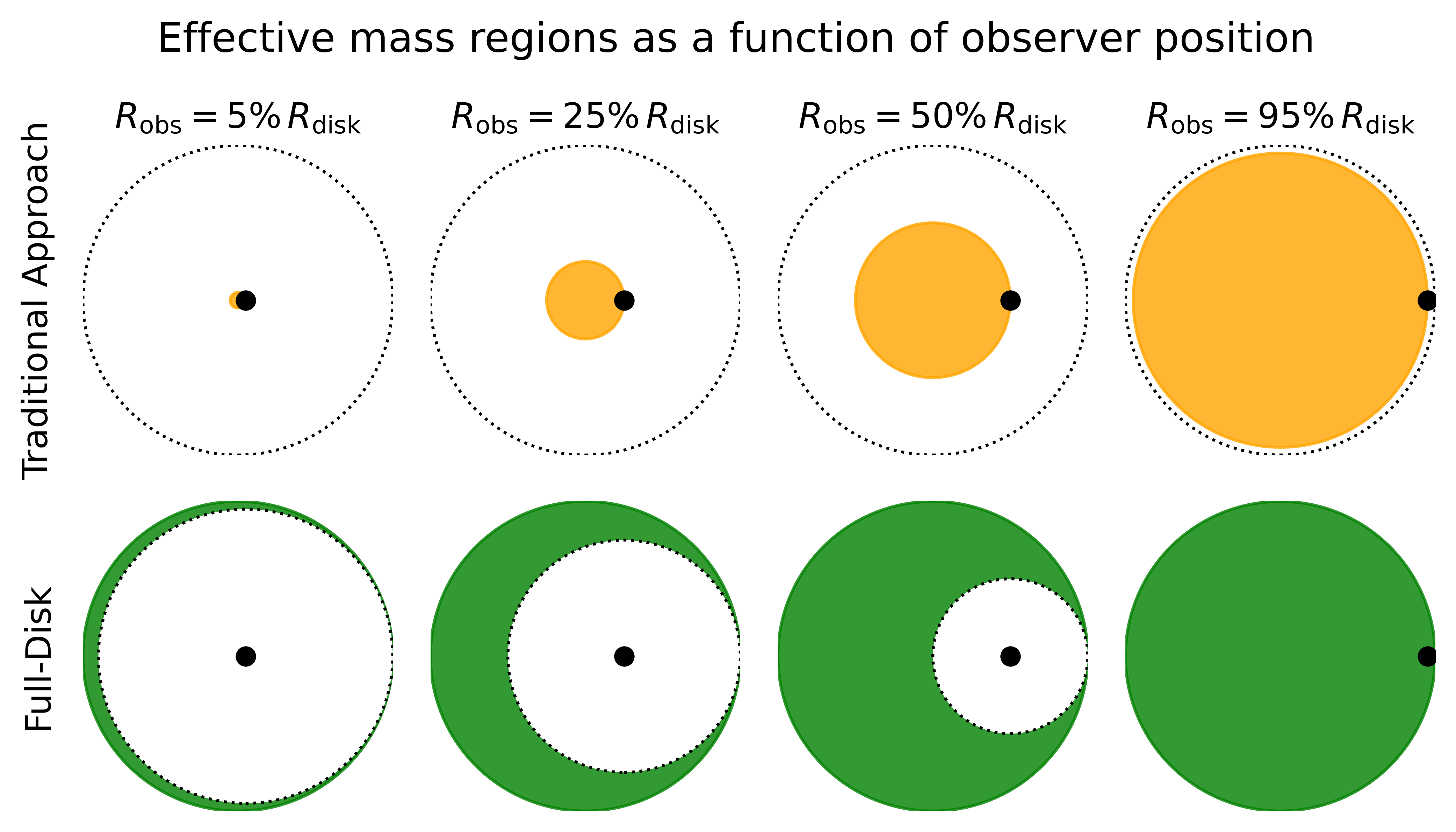}
    \caption{
    Illustration of contributing regions in a simplified disk geometry. The enclosed-mass approximation considers only the central region (top, orange), while a full-disk perspective highlights the contribution of material outside the observer’s radius (bottom, green). The excluded region is the largest disk centered on the observer that remains fully inside the galaxy.
    }
    \label{fig:areas_comparison}
\end{figure}

The remaining region, which contributes most effectively to the net inward acceleration in this simplified construction, forms a crescent-shaped (lune-like) region located opposite the observer. This geometric configuration is illustrated in Figure~\ref{fig:areas_comparison}.

In this simplified construction, the radius of the locally symmetric (excluded) region is given by

\begin{equation}
R_{\mathrm{excl}} = R_{\mathrm{total}} - R_{\mathrm{obs}},
\end{equation}

and the corresponding effective contributing area can be written as

\begin{equation}
A_{\mathrm{eff}} = \pi R_{\mathrm{total}}^2 - \pi (R_{\mathrm{total}} - R_{\mathrm{obs}})^2.
\end{equation}

Expanding this expression yields

\begin{equation}
A_{\mathrm{eff}} = \pi R_{\mathrm{obs}} \left(2 R_{\mathrm{total}} - R_{\mathrm{obs}} \right).
\end{equation}

The ratio between the enclosed-mass approximation and this effective area is therefore

\begin{equation}
\frac{A_{\mathrm{trad}}}{A_{\mathrm{eff}}}
=
\frac{R_{\mathrm{obs}}}{2 R_{\mathrm{total}} - R_{\mathrm{obs}}}.
\end{equation}

\begin{figure}
    \centering
    \includegraphics[width=0.5\textwidth]{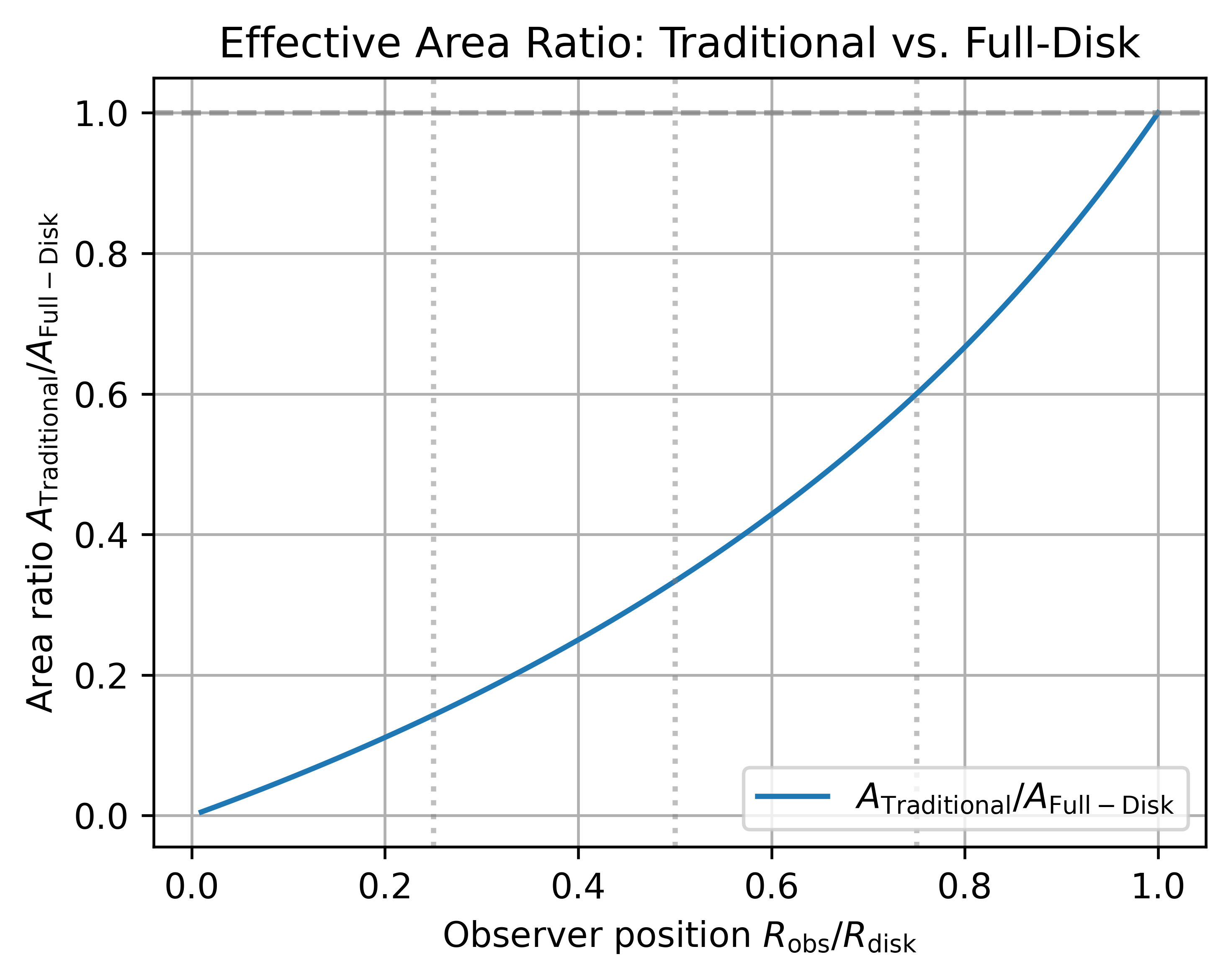}
    \caption{
    Ratio between the enclosed-mass area and the effective contributing area in the simplified disk geometry as a function of observer radius. The difference between both approaches is more pronounced at small radii and decreases toward the disk edge.
    }
    \label{fig:area_ratio}
\end{figure}

The behavior of this ratio is shown in Figure~\ref{fig:area_ratio}.

Although real galaxies do not have uniform surface density, this simplified example highlights the geometric origin of the effect. In more realistic disk models, the same qualitative behavior persists: the contribution of material exterior to the observation radius can remain significant and should be accounted for when evaluating the gravitational field.

\section{The Lost and Found (LF) Model}
\label{sec:lfmodel}

\subsection{Conceptual Framework}

The analysis presented in Section~\ref{sec:traditional} illustrates that enclosed-mass descriptions provide only an approximate representation of disk geometries. In flattened geometries, mass elements located outside the observer's radius do not generally cancel and can therefore contribute to the gravitational field.

Motivated by this observation, we introduce the \textit{Lost and Found} (LF) model. The model combines a geometrically consistent evaluation of the gravitational field through full-disk integration with a two-component parametrization of the disk surface density, allowing both the global disk contribution and the radial mass structure to be treated self-consistently.

In this framework, the gravitational field at any point in the disk is determined by the superposition of contributions from all mass elements. Rather than assuming that only the enclosed mass contributes, the LF model consistently accounts for both interior and exterior regions of the disk. The surface density is represented here through a two-component exponential profile, allowing the model to capture both centrally concentrated and extended mass distributions.

\subsection{Mathematical Formulation}

We model the surface mass density of the galactic disk using a two-component exponential profile:
\begin{equation}
\Sigma(r) = \Sigma_1 e^{-r/r_1} + \Sigma_2 e^{-r/r_2},
\end{equation}
where $\Sigma_1$ and $\Sigma_2$ are the central surface densities of the two components, and $r_1$ and $r_2$ are their corresponding scale lengths.

For clarity and consistency, we define the inner and outer scale lengths as
\begin{equation}
r_{\rm inner} \equiv \min(r_1, r_2), \qquad
r_{\rm outer} \equiv \max(r_1, r_2),
\end{equation}
such that $r_{\rm inner} \leq r_{\rm outer}$ by construction. This convention is purely organizational and does not affect the fitting procedure, but facilitates a consistent interpretation of the results in terms of a centrally concentrated component and an extended disk.

This parametrization provides sufficient flexibility to represent a broad range of galactic mass distributions, including both centrally concentrated and extended structures. In particular, it accommodates deviations from a single exponential profile that are commonly observed in spiral galaxies \citep{freeman1970disks}.

Rather than treating $\Sigma_1$ and $\Sigma_2$ as independent free amplitudes, the model is more conveniently expressed in terms of the total LF-inferred disk mass $M_{\rm LF}$ and a fractional mass parameter $f$, such that
\begin{equation}
M_1 = f\,M_{\rm LF}, \qquad M_2 = (1-f)\,M_{\rm LF},
\end{equation}
where $M_1$ and $M_2$ are the masses associated with the two exponential components as defined in the fitting procedure.

For consistency in the physical interpretation, the corresponding masses are reassigned according to the ordering of the scale lengths, and we define
\begin{equation}
f_{\rm inner} = \frac{M_{\rm inner}}{M_{\rm LF}},
\end{equation}
so that the inner scale length and its associated mass fraction always refer to the same physical component. This definition ensures a one-to-one correspondence between spatial scale and mass contribution.

To represent a disk of finite extent while avoiding an artificial sharp edge, the total surface density is smoothly truncated near the outer radius of the model:
\begin{equation}
\Sigma(r)=\left[\Sigma_1 e^{-r/r_1}+\Sigma_2 e^{-r/r_2}\right]\,C(r),
\end{equation}
where the tapering function is
\begin{equation}
C(r)=\frac{1}{1+\exp\left(\frac{r-R_{\rm cut}}{\Delta}\right)}.
\end{equation}

Here, $R_{\rm cut}$ is the truncation radius and $\Delta$ controls the smoothness of the transition. This tapering ensures that the disk mass remains finite and avoids numerical artifacts associated with a hard outer cutoff.

In the numerical implementation, the surface density is renormalized after the tapering is applied so that the total integrated mass of the final disk model remains equal to $M_{\rm LF}$. Thus, $M_{\rm LF}$ corresponds to the total mass of the truncated model actually used in the gravitational calculation.

In practice, the truncation radius is not treated as an independent free parameter, but is fixed as a fraction of the fitted outer disk extent:
\begin{equation}
R_{\rm cut}=0.95\,R_{\max},
\end{equation}
where $R_{\max}$ is the effective outer radius of the modeled disk. The free structural parameters of the LF model are therefore $(M_{\rm LF}, f, r_1, r_2, R_{\max})$, while the reordered quantities $(r_{\rm inner}, r_{\rm outer}, f_{\rm inner})$ are used for physical interpretation.

The gravitational acceleration at a position $\vec{r}_{\mathrm{obs}}$ in the disk is computed as the vector sum of contributions from all mass elements:
\begin{equation}
\vec{a}(\vec{r}_{\mathrm{obs}}) = -G \iint_{\mathrm{disk}} 
\frac{\Sigma(r') \left(\vec{r}_{\mathrm{obs}} - \vec{r}'\right)}
{\left|\vec{r}_{\mathrm{obs}} - \vec{r}'\right|^3}
\, d^2 r'.
\end{equation}

This expression evaluates the gravitational field generated by a continuous mass distribution in the plane of the disk without assuming spherical symmetry, following the general class of full-disk Newtonian treatments explored in previous studies \citep{casertano1983rotation,jalocha2010,feng2012mass}.

The rotational velocity is then obtained from the radial component of the acceleration:
\begin{equation}
v(R) = \sqrt{R\, a_R(R)},
\end{equation}
where $a_R(R)$ denotes the inward radial component of the acceleration.

This formulation naturally includes the gravitational contribution of both interior and exterior disk mass, providing a Newtonian full-disk description of the gravitational field based on the adopted surface-density model.

In the following section, we describe the numerical implementation used to evaluate this integral and to fit the resulting rotation curves to observational data.

\section{Numerical Methodology}
\label{sec:methodology}

\subsection{Disk Discretization}

The gravitational integral is evaluated numerically by discretizing the galactic disk into concentric rings and angular sectors. The disk is divided into $N_r$ radial bins and $N_\theta$ azimuthal bins, forming a two-dimensional grid in polar coordinates.

Each grid element, located at $\vec{r}' = (r', \theta')$, is assigned a surface density $\Sigma(r')$ and an area element
\begin{equation}
dA = r' \, dr' \, d\theta'.
\end{equation}

The associated mass element is
\begin{equation}
dm_{i,j} = \Sigma(r'_i)\, r'_i\, \Delta r\, \Delta \theta,
\end{equation}
where $\Delta r$ and $\Delta \theta$ are the radial and angular bin sizes, respectively.

After applying the smooth outer tapering, the surface density is renormalized so that the total integrated mass of the discretized disk remains equal to the fitted value $M_{\rm LF}$. The total LF-inferred mass is therefore recovered through
\begin{equation}
M_{\rm LF} = \sum_{i,j} dm_{i,j}.
\end{equation}

\subsection{Acceleration Calculation}

The gravitational acceleration at a given observer position $\vec{r}_{\mathrm{obs}}$ is computed as
\begin{equation}
\vec{a}(\vec{r}_{\mathrm{obs}}) = -G \sum_{i,j}
\frac{dm_{i,j}\,(\vec{r}_{\mathrm{obs}} - \vec{r}'_{i,j})}
{|\vec{r}_{\mathrm{obs}} - \vec{r}'_{i,j}|^3}.
\end{equation}

The radial component $a_R(R)$ is obtained by projection, where $a_R(R)$ denotes the inward radial acceleration. The rotation velocity then follows as
\begin{equation}
v(R) = \sqrt{R\, a_R(R)}.
\end{equation}

To avoid numerical divergences when the observer position approaches a grid cell, a softening length $\epsilon$ is introduced, such that
\begin{equation}
|\vec{r}_{\mathrm{obs}} - \vec{r}'|^2 \rightarrow |\vec{r}_{\mathrm{obs}} - \vec{r}'|^2 + \epsilon^2.
\end{equation}

In the numerical implementation, the softening length is chosen to be equal to the radial grid spacing ($\epsilon=\Delta r$). This regularization preserves the large-scale gravitational behavior while preventing singular contributions from nearby cells.

\subsection{Parameter Optimization}

The LF model was fitted to observed rotation curves by optimizing the parameters $(M_{\rm LF}, f, r_1, r_2, R_{\max})$.

For each parameter set, the model rotation curve was computed and interpolated to the observed radii. The fit quality was quantified using a weighted chi-square statistic when observational uncertainties were available,
\begin{equation}
\chi^2 = \sum_{k} \frac{\left[v_{\mathrm{obs}}(R_k) - v_{\mathrm{model}}(R_k)\right]^2}{\sigma_k^2},
\end{equation}
and by the unweighted sum of squared residuals otherwise.

The optimization was performed in two stages. First, an adaptive grid of $M_{\rm LF}$ values was constructed around an initial dynamical mass estimate based on the outer rotation velocity. For each trial mass, the remaining parameters were optimized using the derivative-free Nelder--Mead simplex algorithm \citep{nelder1965simplex,lagarias1998convergence}, which is well suited for nonlinear optimization problems involving multiple coupled parameters. A second, finer scan around the best solution was then used to refine the optimum and estimate uncertainties using the criterion
\begin{equation}
\Delta \chi^2 \leq 1.
\end{equation}

This hybrid approach improves robustness against local minima and ensures stable convergence across galaxies with different morphologies.

\subsection{Validation Tests}

Validation checks using simplified mass distributions confirmed that the numerical implementation reproduces the expected behavior in idealized configurations. In particular, for approximately uniform inner density profiles, the resulting rotation curves show the expected near-linear rise in the central region, consistent with solid-body-like behavior.

At larger radii, the contribution of extended disk mass naturally modifies the outer rotational behavior and can produce a progressive flattening of the rotation curve, depending on the adopted surface density profile.

Convergence tests show that the results become stable beyond a certain resolution threshold. The adopted softening length scales with the radial grid resolution ($\epsilon=\Delta r$), so that its influence decreases as the numerical mesh is refined.
Typical simulations used $N_r \sim 200$ and $N_\theta \sim 400$, providing a balance between accuracy and computational cost.
Further increases in resolution were found to produce negligible changes in the resulting rotation curves.

\section{Results}
\label{sec:results}

\subsection{Galaxy Sample}

The LF model was applied to a heterogeneous sample of spiral galaxies with high-quality rotation curve data. 
The full list of galaxies analyzed is provided in Table~\ref{tab:results_summary}.

The sample includes both classical systems extensively studied in the literature, particularly in \cite{begeman1991extended}, as well as additional galaxies drawn from more recent compilations such as the SPARC database \citep{lelli2016sparc}, which assembles high-quality rotation curves and associated photometric and H\,{\sc i} measurements from the original observational literature. 
The combined dataset spans a wide range of masses, surface brightness profiles, and rotation curve morphologies, allowing us to test the robustness and general applicability of the LF model across different galactic environments.

For M33, which is not included in the SPARC database, the baryonic mass was adopted from the literature using representative values for the stellar and gas components. We use
\begin{equation}
M_{\rm bar} \approx 8.0\times10^9\,M_\odot,
\end{equation}
consistent with studies of the baryonic distribution and gas content of M33 \citep{Corbelli2014,Gratier2017,Corbelli2003}.

\subsection{Rotation Curves Across the Sample}

\begin{figure*}
    \centering
    \includegraphics[width=\textwidth]{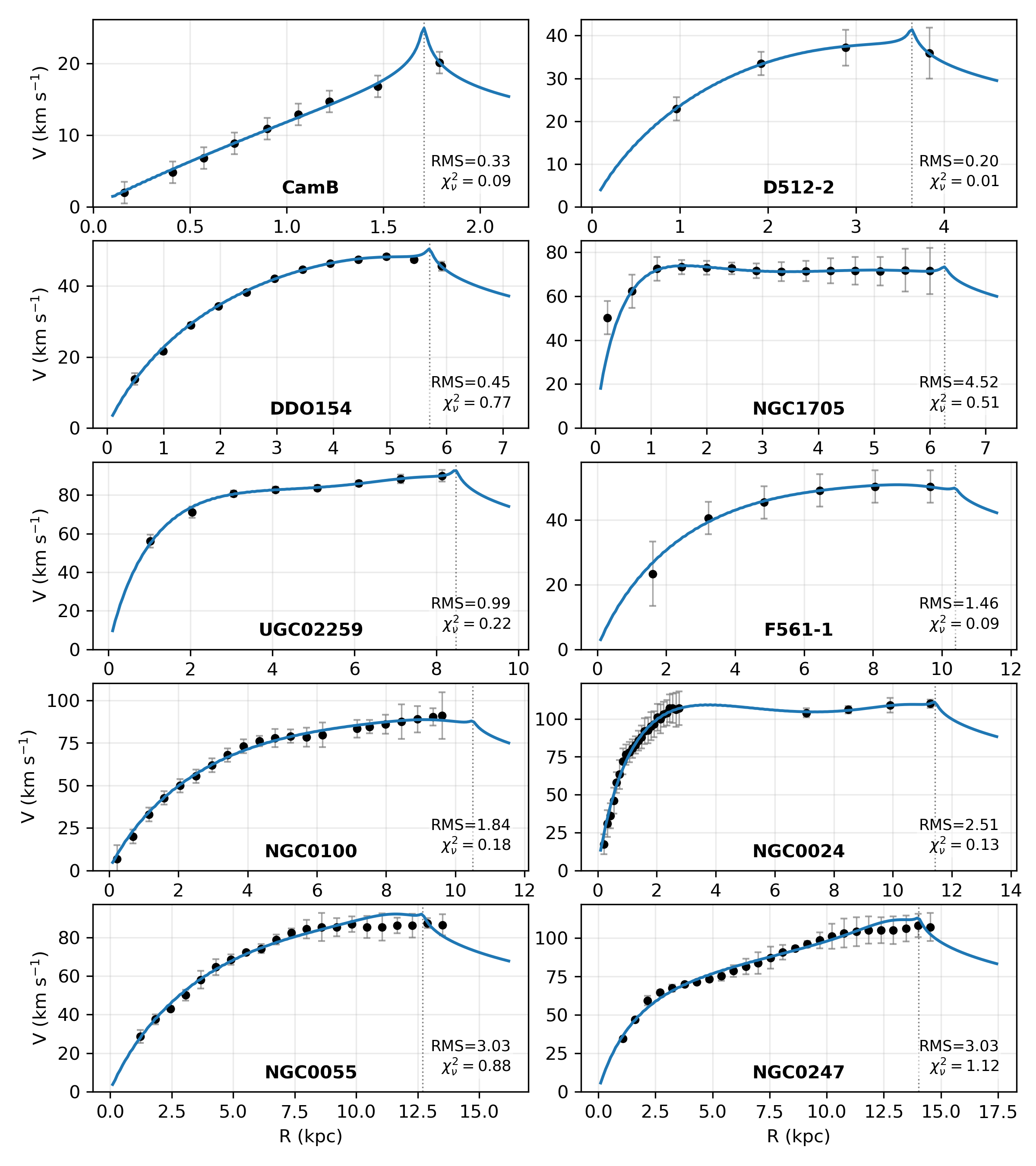}
    \caption{
    Rotation curves for the small- and intermediate-size galaxies in the sample. Observed velocities (points with error bars) are compared with the LF model best-fit curves (solid lines). The galaxies are ordered approximately by increasing radial extent.
    }
    \label{fig:all_curves_small_medium}
\end{figure*}

\begin{figure*}
    \centering
    \includegraphics[width=\textwidth]{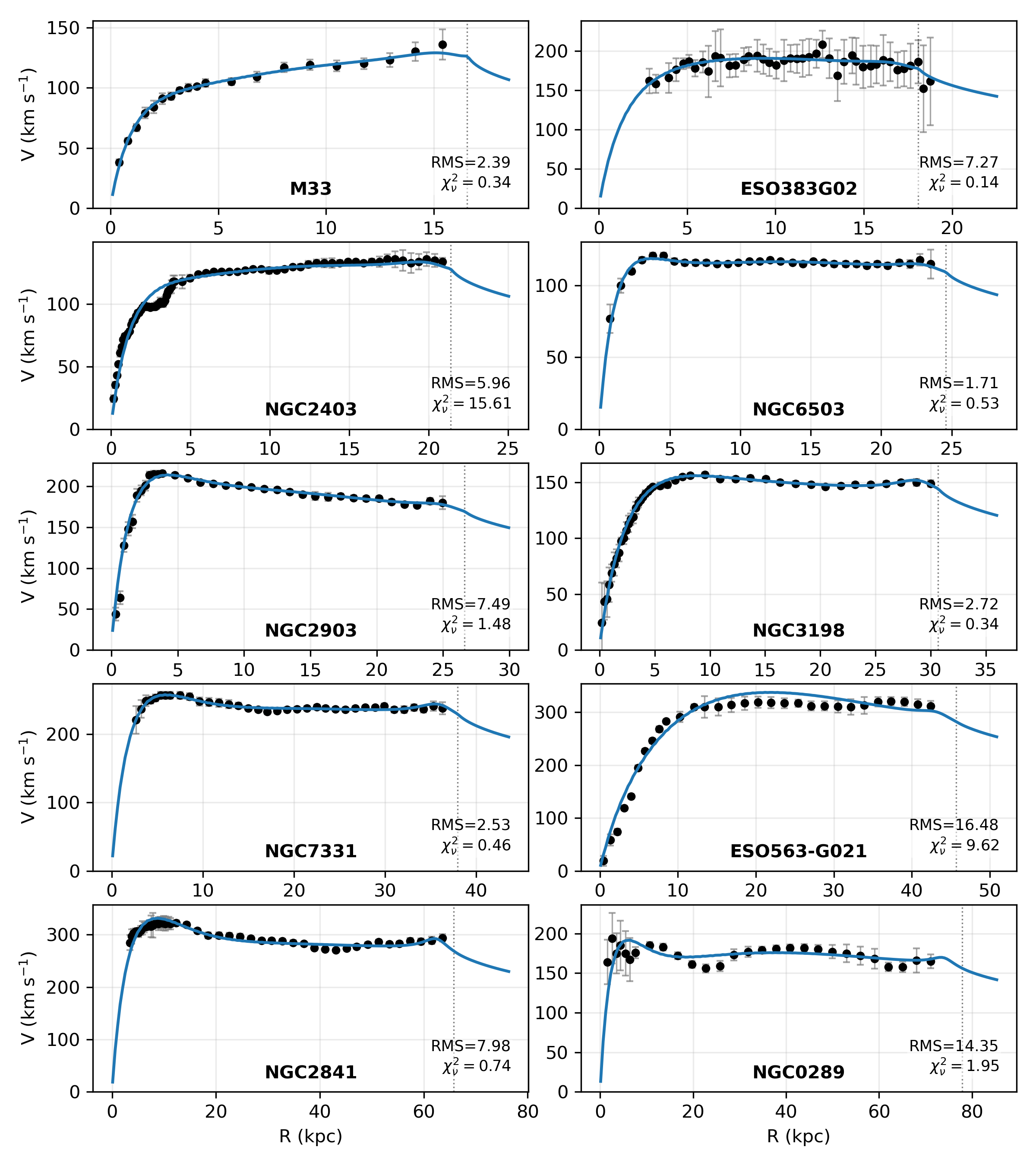}
    \caption{
    Rotation curves for the largest galaxies in the sample. Observed velocities (points with error bars) are compared with the LF model best-fit curves (solid lines). The galaxies are ordered approximately by increasing radial extent.
    }
    \label{fig:all_curves_large}
\end{figure*}

Figures~\ref{fig:all_curves_small_medium} and \ref{fig:all_curves_large} show the observed rotation curves and the corresponding LF model fits for the galaxy sample, ordered approximately by increasing radial extent. 
Observed velocities are shown as points with uncertainties, while the LF best-fit curves are shown as solid lines.

Across the sample, the LF model reproduces the main features of the observed rotation curves, including the inner rising regions and, where present, the transition toward flatter outer behavior. 
The agreement is observed over a broad range of galaxy sizes and kinematic morphologies.

When the galaxies are ordered by increasing $R_{\max}$, a systematic morphological progression becomes apparent. 
Within the present sample, the smallest systems tend to be dominated by a nearly linear inner rise, whereas progressively larger systems reveal increasing curvature, turnover, and extended flatter outer regions. 
This trend suggests a systematic variation in rotation-curve morphology with increasing radial extent.

\subsection{Fit Quality}

The quality of the fits was quantified using the coefficient of determination ($R^2$), the root-mean-square (RMS) residual, and the reduced chi-square when observational uncertainties were available:
\begin{equation}
\mathrm{RMS} = \sqrt{\frac{1}{N} \sum_i \left(v_{\mathrm{obs},i} - v_{\mathrm{model},i}\right)^2},
\end{equation}
\begin{equation}
R^2 = 1 - \frac{\sum_i \left(v_{\mathrm{obs},i} - v_{\mathrm{model},i}\right)^2}{\sum_i \left(v_{\mathrm{obs},i} - \bar{v}_{\mathrm{obs}}\right)^2},
\end{equation}
\begin{equation}
\chi_\nu^2 = \frac{1}{N-p}\sum_i \left(\frac{v_{\mathrm{obs},i}-v_{\mathrm{model},i}}{\sigma_i}\right)^2,
\end{equation}
where $N$ is the number of observed data points, $p$ is the number of free parameters, and $\sigma_i$ are the observational velocity uncertainties.

For most galaxies in the sample, the LF model achieves values of $R^2$ close to unity, indicating excellent agreement with the observed rotation curves. 
Several galaxies, such as NGC 3198, NGC 0024, and UGC 02259, show $R^2 > 0.99$. 
The RMS residuals are typically of the order of a few km\,s$^{-1}$, indicating that the deviations between model and observations remain small in absolute terms.

A small number of galaxies exhibit lower or even formally negative $R^2$ values despite visually reasonable fits. 
This occurs when the observed velocity profile contains strong local fluctuations or large point-to-point scatter, especially in the inner regions, making $R^2$ overly sensitive to deviations relative to the sample mean. 
In such cases, the RMS residual and the reduced chi-square, when available, provide more robust indicators of fit quality than $R^2$ alone.

While $R^2$ provides a convenient measure of visual agreement, $\chi_\nu^2$ is more physically meaningful because it incorporates the observational uncertainties. 
In this sense, some galaxies with modest or even negative $R^2$ values still exhibit statistically acceptable fits, while others with visually smooth agreement may yield larger $\chi_\nu^2$ due to small error bars or localized fluctuations in the observed rotation curve.

\subsection{Radial Extent Comparison}

A useful consistency test is provided by comparing the LF parameter $R_{\max}$ with the maximum observed radial extent of the rotation curve, $R_{\max}^{\rm obs}$, obtained directly from the kinematic data.

We find that these quantities are generally in close agreement:
\begin{equation}
\frac{R_{\max}}{R_{\max}^{\rm obs}} \approx 1,
\end{equation}
with deviations typically within about 10\%.

This result indicates that the LF model naturally recovers the observed radial extent of the disks without requiring external tuning. 
It also suggests that the relevant radial scale in the LF framework is not simply the photometric exponential scale length often quoted in the literature, but rather the full observed kinematic extent of the system.

\subsection{Inner Scale Length and Rotation Curve Transition}

To investigate the physical meaning of the inner LF scale length, we compared the parameter $r_{\rm inner}$,
as defined in Section~\ref{sec:methodology}, with a characteristic turnover radius $r_{\rm turn}$ estimated directly from the observed rotation curves.

The turnover radius was defined as the first radius at which the smoothed local slope of the rotation curve decreases below a fixed fraction of the initial rising slope. This definition provides a simple and robust estimate of the transition between the inner rising regime and the outer, more gradually varying region.

Across the galaxy sample, we find that the ratio $r_{\rm turn}/r_{\rm inner}$ has a typical value of order $\sim 2$, with a median of $\sim 1.9$ and moderate scatter. The 16th–84th percentile range spans approximately $1.3$ to $2.5$, indicating that the transition typically occurs at a few inner scale lengths.

\begin{figure}
    \centering
    \includegraphics[width=0.48\textwidth]{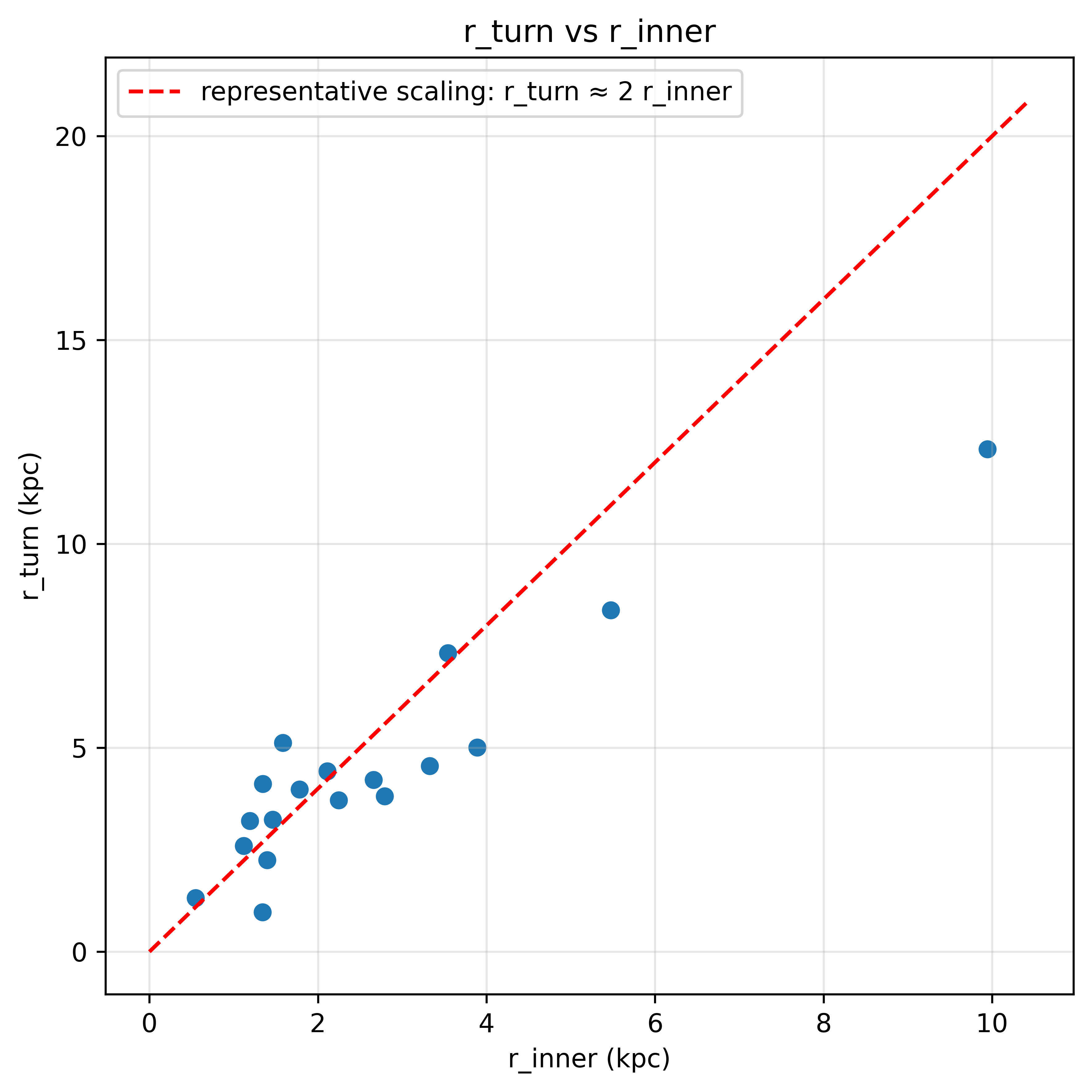}
    \caption{
    Relation between the turnover radius $r_{\rm turn}$ and the inner LF scale length $r_{\rm inner}$. The dashed line indicates the representative scaling $r_{\rm turn} \approx 2\,r_{\rm inner}$, consistent with the median ratio observed across the sample. The scatter reflects variations in the transition between inner and outer dynamical regimes.
    }
    \label{fig:rturn_vs_rinner}
\end{figure}

The relation between $r_{\rm turn}$ and $r_{\rm inner}$ is illustrated in Figure~\ref{fig:rturn_vs_rinner}. The data show a clear positive correlation, with most galaxies lying close to a representative scaling of $r_{\rm turn} \approx 2\,r_{\rm inner}$. Galaxies for which no clear turnover radius could be identified were excluded from the statistical analysis.

This result indicates that the inner exponential component defines a region where the rotation curve exhibits approximately solid-body-like behavior, with velocity increasing nearly linearly with radius. The transition to the flatter regime occurs at several inner scale lengths, reflecting the growing contribution of the extended disk component.

Although the LF model is not intended as a formal photometric decomposition, this behavior suggests that the inner component provides a dynamically meaningful representation of the centrally concentrated mass distribution, while the outer component governs the extended disk.
\subsection{Mass Comparison}

A useful comparison in rotation-curve analyses is how the mass required to reproduce the observed kinematics relates to both the baryonic content and the dynamical mass inferred from the outer rotation speed. 
This comparison is particularly relevant in the context of the long-standing discrepancy between luminous matter and rotation-curve amplitudes reported in disk galaxies \cite{rubin1980rotational,bosma1981distribution,begeman1991extended,sofue2001rotation,deblok2008high,lelli2016sparc}.

To evaluate the LF model in this context, we compared the LF-inferred mass $M_{\rm LF}$ with two independent reference masses.

\begin{figure}
    \centering
    \includegraphics[width=0.5\textwidth]{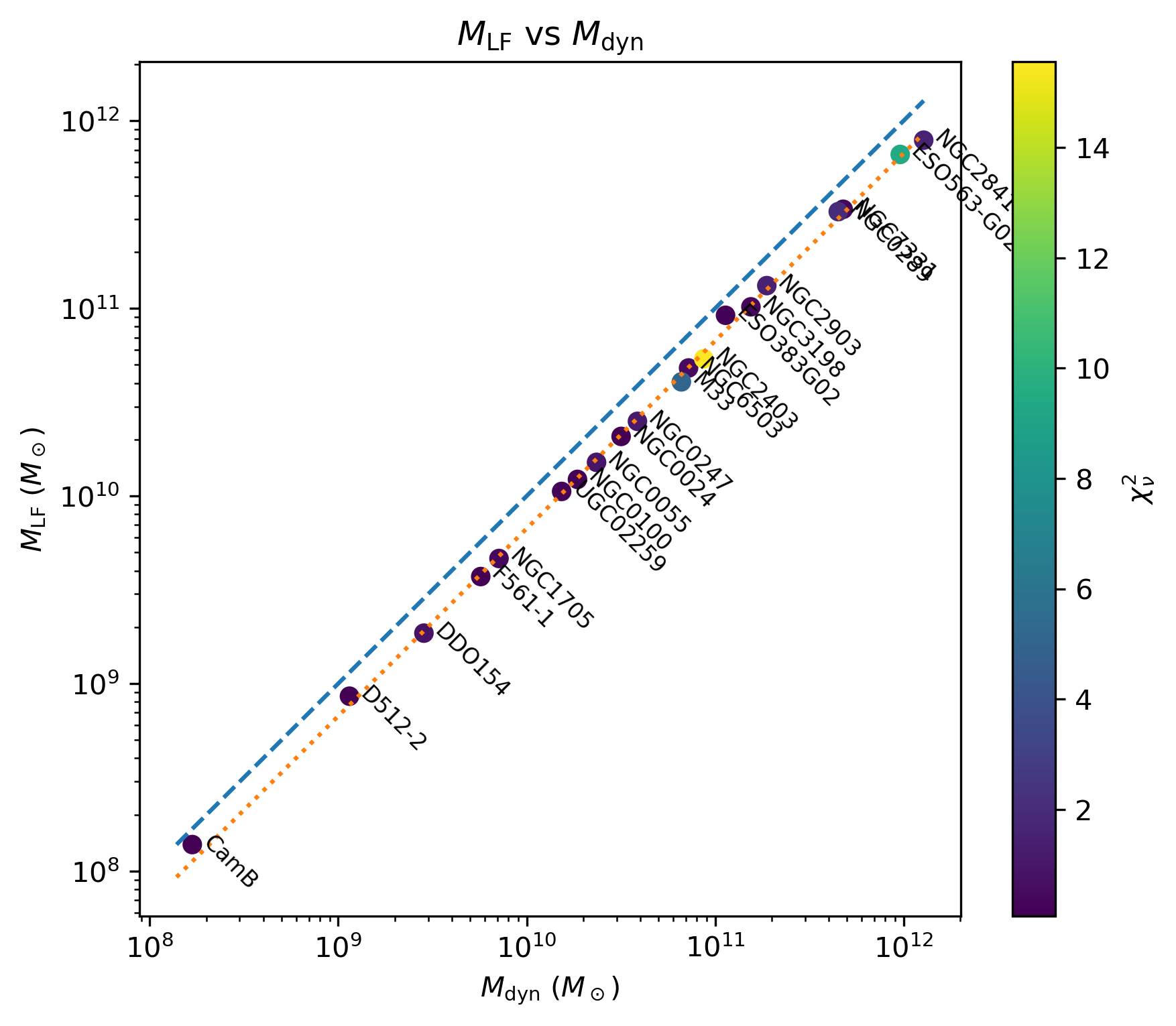}
    \caption{
    Comparison between the LF-inferred mass and the dynamical mass inferred at the outermost observed point of the rotation curve. 
    The dashed line indicates the one-to-one relation, while the dotted line shows the global proportionality trend. 
    Points are color-coded by reduced chi-square.
    }
    \label{fig:MLF_vs_Mdyn}
\end{figure}

First, the baryonic mass was estimated from stellar and gas components as
\begin{equation}
M_{\mathrm{bar}} = \Upsilon_* L_{[3.6]} + 1.33\,M_{\mathrm{HI}},
\end{equation}
where $\Upsilon_*=0.5$ was adopted for the stellar mass-to-light ratio in the 3.6\,$\mu$m band, following the standard assumptions used in SPARC-based analyses \cite{lelli2016sparc,mcgaugh2016radial}.

Second, the dynamical mass was estimated at the outermost observed point of each rotation curve as
\begin{equation}
M_{\mathrm{dyn}}(R_{\max}^{\mathrm{obs}})=\frac{v(R_{\max}^{\mathrm{obs}})^2\,R_{\max}^{\mathrm{obs}}}{G},
\end{equation}
where $v(R_{\max}^{\mathrm{obs}})$ is the observed rotation velocity at the largest measured radius.
This quantity represents the conventional dynamical mass estimate within the observed extent of the rotation curve.
It should be distinguished from virial halo masses or model-dependent dark matter halo masses often quoted in the literature, which generally refer to larger extrapolated scales.

Figure~\ref{fig:MLF_vs_Mdyn} compares the LF-inferred mass with the dynamical mass. 
A consistent relation is observed across the sample. Using the total masses, we obtain
\begin{equation}
\frac{\sum M_{\mathrm{LF}}}{\sum M_{\mathrm{dyn}}}=0.671,
\end{equation}
while the average of galaxy-by-galaxy ratios yields
\begin{equation}
\left\langle \frac{M_{\mathrm{LF}}}{M_{\mathrm{dyn}}} \right\rangle
=0.681\pm0.013,
\end{equation}
with a standard deviation of 0.058 and a median value of 0.658.

A linear fit constrained through the origin gives
\begin{equation}
M_{\mathrm{LF}} = 0.654\,M_{\mathrm{dyn}},
\end{equation}
while the logarithmic fit yields
\begin{equation}
\log M_{\mathrm{LF}} = 0.989\,\log M_{\mathrm{dyn}} - 0.054,
\end{equation}
or equivalently,
\begin{equation}
M_{\mathrm{LF}} \approx 0.883\,M_{\mathrm{dyn}}^{0.989}.
\end{equation}

The logarithmic slope, being very close to unity, indicates that the LF-inferred mass scales almost linearly with the dynamical mass over the full sample. 
Thus, the main difference between the two quantities is not a change in scaling with galaxy mass, but rather a systematic offset in normalization.

To express this trend more compactly, we define the LF mass reduction factor as
\begin{equation}
\eta_{\rm LF} \equiv \frac{M_{\rm LF}}{M_{\rm dyn}},
\end{equation}
where $M_{\rm LF}$ is the LF-inferred mass and $M_{\rm dyn}$ is the standard dynamical mass estimate. Across the galaxy sample analyzed here, this ratio clusters around
\begin{equation}
\eta_{\rm LF} \sim 0.67.
\end{equation}
This behavior is reminiscent of the nearly constant "galactic rotation number" reported in independent thin-disk formulations (e.g., \cite{feng2012mass}), where the total mass is expressed as a proportional scaling of $V^2 R / G$.
\begin{figure}
    \centering
    \includegraphics[width=0.5\textwidth]{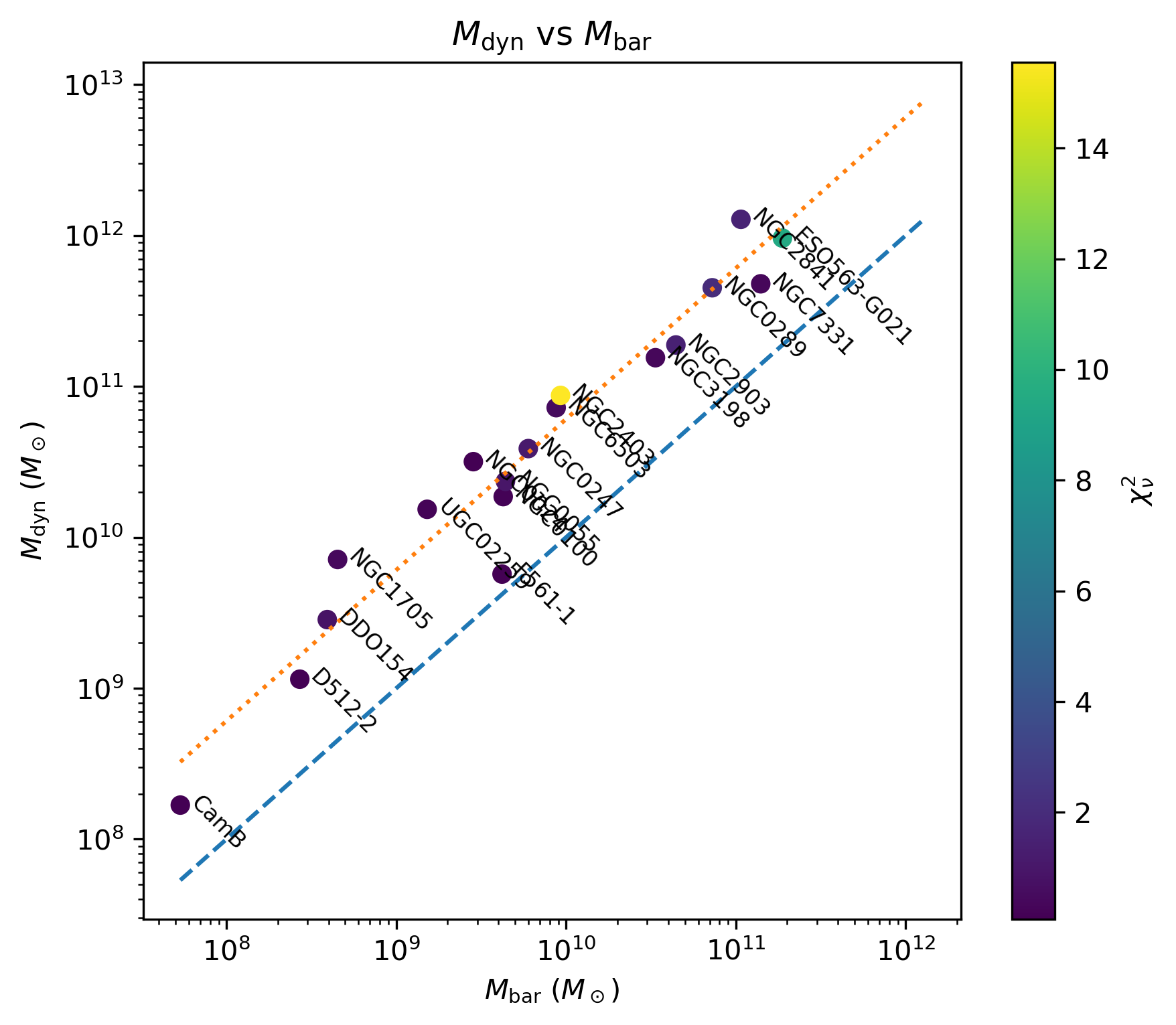}
    \caption{
    Comparison between the dynamical mass and the baryonic mass. 
    The dashed line indicates the one-to-one relation, while the dotted line shows the global proportionality trend. 
    Points are color-coded by reduced chi-square.
    }
    \label{fig:Mdyn_vs_Mbar}
\end{figure}

Figure~\ref{fig:Mdyn_vs_Mbar} compares the dynamical mass with the baryonic mass. 
As expected from classical rotation-curve studies, the dynamical mass lies well above the directly estimated baryonic content for most galaxies. 
Using the total masses, we obtain
\begin{equation}
\frac{\sum M_{\mathrm{dyn}}}{\sum M_{\mathrm{bar}}}=6.075,
\end{equation}
while the galaxy-by-galaxy average gives
\begin{equation}
\left\langle \frac{M_{\mathrm{dyn}}}{M_{\mathrm{bar}}} \right\rangle
=6.80\pm0.87,
\end{equation}
with a standard deviation of 3.68 and a median value of 5.77.

A linear fit constrained through the origin gives
\begin{equation}
M_{\mathrm{dyn}} = 5.77\,M_{\mathrm{bar}},
\end{equation}
while the logarithmic fit yields
\begin{equation}
\log M_{\mathrm{dyn}} = 0.995\,\log M_{\mathrm{bar}} + 0.821,
\end{equation}
or equivalently,
\begin{equation}
M_{\mathrm{dyn}} \approx 6.63\,M_{\mathrm{bar}}^{0.995}.
\end{equation}

The near-unity logarithmic slope indicates that the conventional dynamical mass also scales almost linearly with baryonic mass across the sample, but with a substantially larger normalization than the LF relation discussed above.

\begin{figure}
    \centering
    \includegraphics[width=0.5\textwidth]{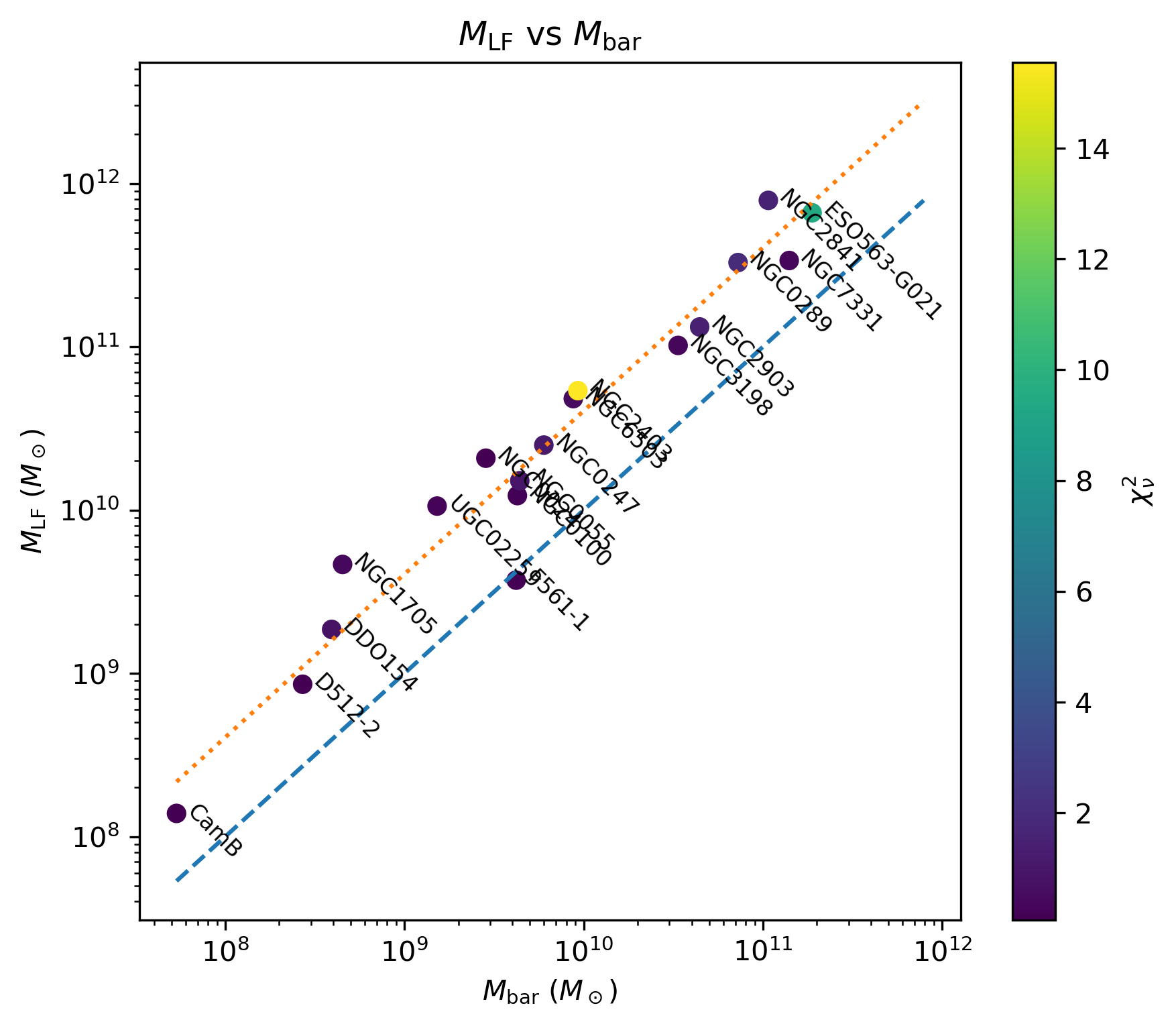}
    \caption{
    Comparison between the LF-inferred mass and the baryonic mass estimated from stellar and H\,{\sc i} data. 
    The dashed line indicates the one-to-one relation, while the dotted line shows the global proportionality trend. 
    Points are color-coded by reduced chi-square.
    }
    \label{fig:MLF_vs_Mbar}
\end{figure}

Figure~\ref{fig:MLF_vs_Mbar} compares the LF-inferred mass with the baryonic mass. 
In contrast to the comparison with $M_{\rm dyn}$, the LF-inferred mass systematically exceeds the baryonic estimate, although by a smaller factor than the standard dynamical mass does. 
Using the ratio of total masses, we find
\begin{equation}
\frac{\sum M_{\mathrm{LF}}}{\sum M_{\mathrm{bar}}}=4.05,
\end{equation}
while the average galaxy-by-galaxy ratio is
\begin{equation}
\left\langle \frac{M_{\mathrm{LF}}}{M_{\mathrm{bar}}} \right\rangle
=4.53\pm0.54,
\end{equation}
with a standard deviation of 2.31 and a median value of 3.85.

A linear fit constrained through the origin gives
\begin{equation}
M_{\mathrm{LF}} = 3.87\,M_{\mathrm{bar}},
\end{equation}
while the logarithmic fit yields
\begin{equation}
\log M_{\mathrm{LF}} = 0.985\,\log M_{\mathrm{bar}} + 0.748,
\end{equation}
or equivalently,
\begin{equation}
M_{\mathrm{LF}} \approx 5.60\,M_{\mathrm{bar}}^{0.985}.
\end{equation}

\begin{table*}
\centering
\caption{
Summary of LF model fitted parameters. 
Here, $N$ is the number of observed rotation-curve points; 
$M_{\rm LF}$ is the total LF-inferred mass; 
$M_{\rm bar}$ is the baryonic mass; 
$M_{\rm dyn}$ is the conventional dynamical mass estimated at the outermost observed radius;
$f_{\rm inner}$ is the mass fraction associated with the inner exponential component 
($M_{\rm inner}=f_{\rm inner} M_{\rm LF}$); 
$r_{\rm inner}$ and $r_{\rm outer}$ are the scale lengths of the inner and outer components, defined such that $r_{\rm inner} \leq r_{\rm outer}$; 
$R_{\max}^{\rm obs}$ is the maximum observed radius; 
$R_{\max}$ is the best-fit truncation radius; 
$v(R_{\max}^{\rm obs})$ is the observed velocity at the outermost point; 
and $\chi_\nu^2$ is the reduced chi-square of the fit.
}
\label{tab:results_summary}
\footnotesize
\setlength{\tabcolsep}{3.5pt}
\begin{tabular}{lccccccccccc}
\toprule
Galaxy & $N$ & $M_{\rm LF}$ & $M_{\rm bar}$ & $M_{\rm dyn}$ & $f_{\rm inner}$ & $r_{\rm inner}$ & $r_{\rm outer}$ & $R_{\max}^{\rm obs}$ & $R_{\max}$ & $v(R_{\max}^{\rm obs})$ & $\chi_\nu^2$ \\
&  & ($M_\odot$) & ($M_\odot$) & ($M_\odot$) &  & (kpc) & (kpc) & (kpc) & (kpc) & (km/s) &  \\
\midrule
CamB         & 9  & $1.39\times10^8$   & $5.35\times10^7$   & $1.68\times10^8$   & 0.010 & 15.00 & 29.42 & 1.79  & 2.05  & 20.1 & 0.071 \\
D512-2       & 4  & $8.57\times10^8$   & $2.70\times10^8$   & $1.15\times10^9$   & 0.959 & 1.69  & 6.30  & 3.83  & 4.30  & 35.9 & 0.106 \\
DDO154       & 12 & $1.86\times10^9$   & $3.92\times10^8$   & $2.85\times10^9$   & 0.972 & 2.60  & 20.10 & 5.92  & 5.71  & 45.5 & 0.848 \\
NGC1705      & 14 & $4.64\times10^9$   & $4.51\times10^8$   & $7.13\times10^9$   & 0.280 & 0.50  & 3.10  & 6.00  & 5.89  & 71.5 & 0.388 \\
F561-1       & 6  & $3.72\times10^9$   & $4.20\times10^9$   & $5.71\times10^9$   & 0.985 & 3.52  & 3.89  & 9.66  & 10.39 & 50.4 & 0.087 \\
UGC02259     & 8  & $1.06\times10^{10}$ & $1.52\times10^9$   & $1.53\times10^{10}$ & 0.309 & 1.27  & 6.03  & 8.14  & 9.05  & 90.0 & 0.219 \\
NGC0100      & 21 & $1.23\times10^{10}$ & $4.26\times10^9$   & $1.86\times10^{10}$ & 0.962 & 3.90  & 60.00 & 9.62  & 11.04 & 91.2 & 0.160 \\
NGC0024      & 29 & $2.08\times10^{10}$ & $2.84\times10^9$   & $3.17\times10^{10}$ & 0.406 & 1.41  & 9.23  & 11.27 & 11.69 & 110  & 0.129 \\
NGC0055      & 21 & $1.51\times10^{10}$ & $4.40\times10^9$   & $2.35\times10^{10}$ & 0.066 & 5.33  & 5.33  & 13.50 & 13.76 & 86.5 & 0.929 \\
NGC0247      & 26 & $2.50\times10^{10}$ & $5.99\times10^9$   & $3.87\times10^{10}$ & 0.075 & 1.66  & 8.27  & 14.54 & 15.18 & 107  & 1.153 \\
M33          & 20 & $4.06\times10^{10}$ & $8.00\times10^9$   & $6.62\times10^{10}$ & 0.102 & 1.21  & 7.34  & 15.40 & 16.94 & 136  & 5.048 \\
ESO383G02    & 42 & $9.19\times10^{10}$ & ---                & $1.14\times10^{11}$ & 0.153 & 1.59  & 5.36  & 18.78 & 18.09 & 161  & 0.136 \\
NGC2403      & 73 & $5.39\times10^{10}$ & $9.28\times10^9$   & $8.71\times10^{10}$ & 0.128 & 1.34  & 6.92  & 20.87 & 20.97 & 134  & 15.553 \\
NGC6503      & 31 & $4.81\times10^{10}$ & $8.74\times10^9$   & $7.23\times10^{10}$ & 0.177 & 1.24  & 7.44  & 23.50 & 24.58 & 115  & 0.507 \\
NGC2903      & 34 & $1.32\times10^{11}$ & $4.43\times10^{10}$ & $1.88\times10^{11}$ & 0.252 & 1.46  & 7.22  & 24.96 & 26.62 & 180  & 1.479 \\
NGC3198      & 43 & $1.02\times10^{11}$ & $3.36\times10^{10}$ & $1.55\times10^{11}$ & 0.297 & 2.79  & 11.62 & 30.00 & 30.69 & 149  & 0.341 \\
NGC7331      & 36 & $3.37\times10^{11}$ & $1.40\times10^{11}$ & $4.78\times10^{11}$ & 0.221 & 2.15  & 13.77 & 36.31 & 39.21 & 238  & 0.363 \\
ESO563-G021  & 30 & $6.62\times10^{11}$ & $1.88\times10^{11}$ & $9.60\times10^{11}$ & 0.865 & 9.42  & 60.00 & 42.41 & 45.69 & 312  & 9.403 \\
NGC2841      & 50 & $7.88\times10^{11}$ & $1.07\times10^{11}$ & $1.28\times10^{12}$ & 0.262 & 3.38  & 22.96 & 63.64 & 65.80 & 294  & 1.530 \\
NGC0289      & 28 & $3.28\times10^{11}$ & $7.26\times10^{10}$ & $4.50\times10^{11}$ & 0.148 & 2.25  & 20.61 & 71.12 & 77.86 & 165  & 1.951 \\
\bottomrule
\end{tabular}
\end{table*}

The logarithmic slope remains close to unity, suggesting that the LF-inferred mass broadly follows the baryonic mass scaling across the sample, albeit with substantially larger scatter than in the comparison with $M_{\rm dyn}$. 
This larger scatter is expected, as baryonic mass estimates are subject to additional systematic uncertainties, including stellar mass-to-light ratios and gas mass determinations.

Taken together, these comparisons show that the LF model occupies an intermediate scaling between the observed baryonic mass and the conventional dynamical mass. 
The required LF mass is systematically larger than the directly observed baryonic content, but also systematically smaller than the dynamical mass inferred from the outer rotation velocity. 
This intermediate scaling emerges robustly across the galaxy sample and appears to be a characteristic feature of the LF framework.

An especially notable case is F561-1, for which the LF-inferred mass is nearly consistent with the estimated baryonic mass.

\subsection{Best-Fit Parameters}

Table~\ref{tab:results_summary} summarizes the fitted structural parameters of the LF model for the galaxy sample, together with the corresponding baryonic and dynamical reference masses.

The fitted parameters show a consistent two-scale structure, with a compact inner component and a more extended outer component, reflecting the need to reproduce both the steep inner rise and the extended outer regions of the observed rotation curves.

The parameter $f_{\rm inner}$ represents the fraction of the total LF-inferred mass associated with the inner exponential component ($M_{\rm inner} = f_{\rm inner} M_{\rm LF}$). Its value varies substantially across the sample, indicating that the LF model does not impose a universal fixed partition between inner and outer structure. Instead, the relative contribution of both components adapts naturally to the observed kinematic profile of each galaxy.

This formulation ensures a direct correspondence between spatial scale and mass contribution, allowing the inner dynamical structure to be characterized consistently across the sample.
\section{Discussion}
\label{sec:discussion}

\subsection{Geometric Interpretation of the LF Correction}

The results obtained with the LF framework indicate that geometric effects associated with the full disk mass distribution can contribute significantly to the observed rotation-curve behavior, including the tendency toward outer flattening.

In conventional analyses, dynamical mass estimates are often based on the relation
\begin{equation}
v^2(R) = \frac{G M(<R)}{R},
\end{equation}
which is exact only under the assumption of spherical symmetry \citep{binney2011galactic}.
For flattened systems, however, mass exterior to the radius of interest can contribute to the gravitational field and is not generally represented by the enclosed-mass approximation.
The LF model shows that when the gravitational field is evaluated through direct integration of the full disk, the outer mass continues to contribute significantly to the acceleration at interior radii \citep{freeman1970disks,casertano1983rotation}.

As a result, the same observed rotation curves can be reproduced with systematically lower inferred masses than those obtained from conventional dynamical mass estimates. Across the present sample, this manifests as a nearly constant mass scaling factor $\eta_{\rm LF} \sim 0.67$. This behavior suggests that part of the inferred mass discrepancy, within the sample analyzed here, may be associated with the geometric assumptions underlying standard mass estimates

The introduction of the reordered parameters $(r_{\rm inner}, r_{\rm outer}, f_{\rm inner})$ allows for a consistent interpretation of the LF model in which spatial scale and mass contribution are directly linked. In this framework, the inner scale length and its associated mass fraction describe a dynamically meaningful central region, while the outer component accounts for the extended disk. The absence of a fixed value of $f_{\rm inner}$ across the sample further highlights the flexibility of the model in adapting to different galactic structures.
This result further indicates that the inner dynamical structure of disk galaxies is not characterized by a universal mass fraction, but rather emerges from the interplay between geometry and the observed rotation profile.
\subsection{Relation to Baryonic Mass and Residual Discrepancy}

Although the LF model reduces the discrepancy relative to $M_{\rm dyn}$, the LF-inferred mass remains systematically larger than the estimated baryonic mass. This residual difference can arise from several sources.

Baryonic mass estimates depend on photometric light profiles, H\,{\sc i} measurements, and assumptions about mass-to-light ratios, gas corrections, and low-surface-brightness outskirts \citep{deblok2008high,lelli2016sparc}. These factors introduce uncertainties that may lead to an underestimation of the total effective gravitating disk mass.

In addition, the LF surface density profile is an effective parametrization. The fitted mass distribution may also absorb part of the contribution from extended or diffuse structures that are not always fully captured in simplified photometric decompositions.

The position of $M_{\rm LF}$ between $M_{\rm bar}$ and $M_{\rm dyn}$ therefore suggests that geometric effects can account for a significant fraction of the inferred discrepancy, even if a residual mass excess remains to be explained \citep{mcgaugh2016radial}.

\subsection{Interpretive Scope of the LF Model}

The LF framework does not introduce new mass components or modify Newtonian gravity. Instead, it revisits how the gravitational field is computed in flattened systems. In this sense, it is distinct from both dark matter halo models and modified-gravity approaches such as MOND.

The LF-inferred surface density should be interpreted as an effective gravitating distribution that reproduces the observed kinematics, rather than a unique reconstruction of the underlying baryonic mass. The adopted two-component exponential profile provides a simplified effective description of the radial mass distribution, but does not capture the full structural complexity of real galaxies.

Accordingly, the present results should be understood as demonstrating the impact of geometric treatment on mass inference, rather than as a definitive decomposition of the underlying baryonic structure.

In this sense, the LF framework is closely related to previous thin-disk approaches that compute the gravitational field through direct integration of the disk mass distribution (e.g., \citealt{jalocha2010, feng2012mass}), while providing a simplified parametrization suitable for systematic fitting across heterogeneous galaxy samples.

\subsection{Observational and Modeling Limitations}

Several limitations of the present analysis should be noted.

First, the use of a two-component exponential profile is a simplified representation of the disk structure. While it reproduces a wide range of rotation-curve morphologies, it is not intended as a unique or fully realistic description of the stellar and gaseous mass distribution.

Second, the model assumes an axisymmetric thin disk. Non-axisymmetric features such as bars, spiral arms, and local asymmetries are not explicitly included.

Third, the baryonic comparison is limited by the availability and consistency of observational data.

Finally, the dynamical mass $M_{\rm dyn}$ used for comparison is derived from the observed velocity at the outermost measured radius.

\subsection{Implications and Future Directions}

The results presented here motivate several directions for further investigation.

Applying the LF framework to larger and more homogeneous galaxy samples will be essential to assess the robustness of the mass reduction factor across different galaxy types.

A particularly important step will be to replace the present effective mass profiles with surface density distributions directly constrained from observed stellar and gas maps.

These extensions will help quantify the extent to which geometric effects alone can account for the apparent mass discrepancy, and how much may still require additional physical ingredients.

\section{Conclusions}
\label{sec:conclusions}

We have shown that a geometrically consistent Newtonian treatment of flattened disk systems can reproduce the main observed features of galactic rotation curves while systematically reducing the inferred total mass relative to conventional dynamical mass estimates.

Across the present sample, the LF framework yields a characteristic mass scaling factor $\eta_{\rm LF} \sim 0.67$, suggesting that part of the inferred mass discrepancy in disk galaxies may be associated with geometric assumptions in standard mass estimates. Although the LF-inferred mass generally remains larger than the directly estimated baryonic mass, the results suggest that geometric effects may account for a substantial fraction of the inferred discrepancy.

These findings motivate a reassessment of how gravitational fields are evaluated in non-spherical systems, and highlight the importance of full-disk treatments when interpreting rotation curves. Future work should test the LF framework using directly observed stellar and gas surface density profiles, as well as larger and more diverse galaxy samples, to further quantify the role of geometry in shaping galactic dynamics.
\section*{Acknowledgements}

The author gratefully acknowledges the academic environment and institutional support provided by the University of New Hampshire, which helped make this work possible. Work at the University of New Hampshire is supported by NASA grants 80NSSC24K1245, 80NSSC21K0463, and 80NSSC23K1057. This research was developed independently and outside the primary scientific scope of those funded activities.

This work made use of the SPARC database \citep{lelli2016sparc}. The author further acknowledges the original observational studies compiled therein, whose high-quality rotation-curve and photometric data form an essential foundation for the present analysis.

\section*{Data Availability}

The data underlying this article are publicly available from the SPARC database \citep{lelli2016sparc}, accessible at \url{http://astroweb.cwru.edu/SPARC/}, together with a small number of additional literature sources cited explicitly in the text. The numerical outputs and derived data products generated in this work are available from the author upon reasonable request.

%==================================================
% Optional appendix
%==================================================
%\appendix

%\section{Optional Appendix Title}
% Paste appendix material here if needed.

%==================================================
%                 BIBLIOGRAPHY
%==================================================
\bibliographystyle{aasjournal}
\bibliography{example}

\end{document}